\documentstyle[12pt,epsf]{article}

\def\lsim{\raise0.3ex\hbox{$<$\kern-0.75em\raise-1.1ex\hbox{$\sim$}}}
\def\gsim{\raise0.3ex\hbox{$>$\kern-0.75em\raise-1.1ex\hbox{$\sim$}}}

\begin{document}

\sloppy
\begin{titlepage}

\begin{flushright}
CERN-TH/96-126\\
HD-THEP-96-15\\
IUHET-333\\
hep-ph/9605288\\
May 10, 1996
\end{flushright}
\begin{centering}
\vfill
{\large\bf Is There a Hot Electroweak Phase Transition at $m_H \gsim m_W$?}

\vspace{0.4cm}
K. Kajantie$^{\rm a,b,}$\footnote{keijo.kajantie@cern.ch},
M. Laine$^{\rm c,}$\footnote{M.Laine@thphys.uni-heidelberg.de},
K. Rummukainen$^{\rm d,}$\footnote{kari@trek.physics.indiana.edu}, and
M. Shaposhnikov$^{\rm a,}$\footnote{mshaposh@nxth04.cern.ch} \\

\vspace{0.4cm}
{\em $^{\rm a}$ Theory Division, CERN, CH-1211 Geneve 23, Switzerland}

\vspace{0.1cm}
{\em $^{\rm b}$ Department of Physics,
P.O.Box 9, 00014 University of Helsinki, Finland}

\vspace{0.1cm}
{\em $^{\rm c}$ Institut f\"ur Theoretische Physik,
Philosophenweg 16,
D-69120 Heidelberg, Germany}

\vspace{0.1cm}
{\em $^{\rm d}$ Indiana University, Department of Physics,
Swain Hall West 117, Bloomington IN 47405, USA}

\vspace{5mm}
{\bf Abstract}

\end{centering}

\vspace{0.3cm}\noindent
We provide non-perturbative evidence for the fact that there is no hot
electroweak phase transition at large Higgs masses, $m_H = 95$, 120
and 180 GeV\@. This means that the line of first order phase
transitions separating the symmetric and broken phases at small $m_H$
has an end point $m_{H,c}$. In the minimal standard electroweak theory
70 GeV $<m_{H,c}<$ 95 GeV and most likely $m_{H,c} \approx 80$ GeV. If
the electroweak theory is weakly coupled and the Higgs boson is found
to be heavier than the critical value (which depends on the theory in
question), cosmological remnants from the electroweak epoch are
improbable.

\vspace{3mm}\noindent
pacs: 11.10.Wx, 11.15.Ha

\vfill \vfill
\noindent
CERN-TH/96-126\\
HD-THEP-96-15\\
IUHET-333
\end{titlepage}

\noindent
The transition between the high temperature symmetric (or confinement)
phase and the low $T$ broken (or Higgs) phase in the standard
electroweak theory or its extensions is known to be of first order for
small values of the Higgs mass $m_H$. This follows from perturbative
studies of the effective potential \cite{effpot} and non-perturbative
lattice Monte Carlo simulations \cite{4dfirst,3dlatt,4dlatt}. In the
region of applicability of the perturbative expansion the strength of
the electroweak phase transition decreases when $m_H$ increases.
However, the nature of the electroweak phase transition at
``large'' Higgs masses, $m_H
{\raise0.3ex\hbox{$>$\kern-0.75em\raise-1.1ex\hbox{$\sim$}}} 
m_W$ remains unclear, since the
perturbative expansion for the description of the phase transition is
useless there. This letter contains the results of the first
non-perturbative MC analysis of the problem for ``large'' Higgs
masses, $m_H=95, 120, 180$ GeV\@.  We shall show that the system behaves
very regularly there, much like water above the critical point. As
there is no distinction between liquid water and vapor, there is no
distinction between the symmetric and broken phases;
there is no long-range order.

In ref. \cite{3dred} it has been shown that in a weakly coupled
electroweak theory and in most of its extensions (supersymmetric or
not) the hot EW phase transition can be described by an
SU(2)$\times$U(1) gauge-Higgs model in three Euclidean
dimensions. (We stress that our study is not applicable for a strongly
coupled Higgs sector, where the perturbative scheme of dimensional
reduction is not valid.)
Since the effects of the U(1) group are perturbative deep in the
Higgs phase and high in the symmetric phase, the presence of the U(1)
factor cannot change the qualitative features of the phase diagram of
this theory. Thus we shall neglect the U(1) factor and work in the
limit $\sin\theta_W =0$. The effective Lagrangian is
\begin{equation}
L  =
\frac{1}{4}G^a_{ij}G^a_{ij}+
(D_i\phi)^{\dagger}(D_i\phi)+
{m}_3^2\phi^{\dagger}\phi+{\lambda}_3
(\phi^{\dagger}\phi)^2 ,
\label{univers}
\end{equation}
where $G^a_{ij}$ is the SU(2) field strength, $\phi$ is the scalar
doublet and $D_i$ is the covariant derivative. The three parameters of
the 3d theory (gauge coupling $g_3^2$, scalar self-coupling
$\lambda_3$ and the scalar mass $m_3^2$) depend on temperature and on
underlying 4d parameters and can be computed perturbatively; the
explicit relations for the MSM are worked out in \cite{3dred} and for
MSSM in \cite{mikko}. The phase structure of the theory
(\ref{univers}) depends on one dimensionless ratio,
$x=\lambda_3/{g_3^2}$, because the dimensionful coupling $g_3^2$ can
be chosen to fix the scale, while the change of the second
dimensionless ratio $y= {m_3^2(g_3^2)}/{g_3^4}$ corresponds to
temperature variation. For $y \gg 1$ (large $T$) the system is in the
strongly coupled symmetric phase, while at $y\ll -1$ (low $T$) the
system is in the weakly coupled Higgs phase. In presenting our results
we will use a more physical set of variables $m_H^*$ and $T^*$ instead
of $x$ and $y$. The parameter $m_H^*$ is the tree-level Higgs mass in
the 4d SU(2)+Higgs theory and $T^*$ is the temperature there. The
exact relationship between $(x,y)$ and $(m_H^*,T^*)$ is given in
eqs.~(2.9--10) of~\cite{3dlatt}.

\begin{figure}[t]
\vspace*{-1.5cm}
\hspace{0.5cm}
\epsfysize=15cm
\centerline{\epsffile{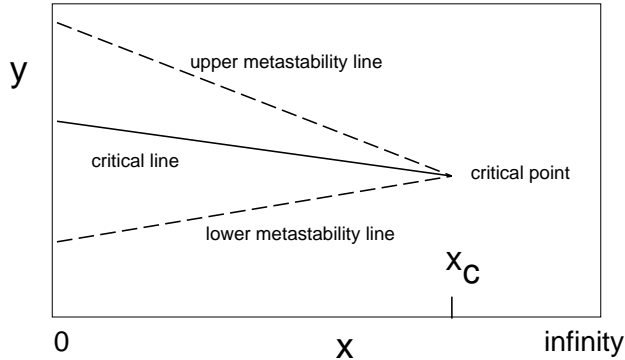}}
\vspace{-9cm}
\caption[a]{\protect
The schematical phase diagram for the SU(2) gauge-Higgs theory.
Solid line is the phase transition and dashed lines indicate
the metastability region.}
\label{phdiagr}
\end{figure}

An essential point in understanding the phase structure of the theory
is the fact that the 3d gauge-Higgs system (as well as the underlying
electroweak theory) does not have a true gauge-invariant order
parameter which can distinguish the symmetric high temperature phase
and low temperature Higgs phase \cite{banks,fradk}. There is no
breaking or restoration of the gauge symmetry across the phase
transition, just because physical observables are always gauge
invariant. The physical spectrum of the corresponding Minkowskian
$(2+1)$ theory in the Higgs phase consists of three massive vector
bosons and one scalar excitation, perfectly mapping to the spectrum of
low lying resonances (three vector bound states of scalar constituent
``quarks'' and one scalar bound state) in the symmetric phase. The
corresponding scalar ($\pi$) and vector ($V$) gauge-invariant
operators are given by $\pi = \phi^{\dagger}\phi$,
$V_j^{0}=i\phi^{\dagger}\stackrel{\leftrightarrow}{D}_j\phi$,
$V_j^{+}=(V_j^{-})^*= 2i\phi^{\dagger}D_j{\tilde\phi}$, where
$\tilde\phi=i\sigma_2\phi^*$.

In lattice non-abelian gauge-Higgs systems with matter in the
fundamental representation and fixed length of the scalar field, the
Higgs (weakly coupled) and symmetric (strongly coupled) phases are
continuously connected \cite{fradk}. This suggests the phase diagram
on the ($x,y$) (Higgs mass-temperature) plane shown in Fig.~1. The
knowledge of the phase diagram and the value of $x_c$ is essential for
cosmological applications. If $x_c = \infty$, the electroweak phase
transition did occur in the early Universe at the electroweak scale
independent of the parameters of the electroweak theory. This means
that substantial deviations from thermal equilibrium took place at
this scale, which might leave some observable remnants such as the
baryon asymmetry of the universe (for a review see~\cite{rs} and
references therein). In the opposite situation of finite $x_c$ the EW
phase transition never took place for a region of parameters of the
underlying theory; in this case it is extremely unlikely that there
are any remnants from the electroweak epoch.

There were up to now no solid results on the phase structure of the
continuum 3d (and, therefore, high temperature) gauge-Higgs theory.
Various arguments in favour and against finite $x_c$ are listed
below.

1. $x_c=\infty$?  The limit $x \rightarrow \infty$ corresponds
formally to $g_3^2=0$, i.e. to the pure scalar model with SU(2)
global symmetry. The latter is known to have a second order phase
transition, suggesting that $x_c=\infty$ in the SU(2) gauge-Higgs
system. The weakness of this argument is revealed when the particle
spectra of the two theories are compared: the pure scalar theory
below the critical point contains massless scalar 
particles -- Goldstone bosons -- 
but the spectrum of the gauge theory contains only
massive modes.

The $\epsilon$-expansion predicts a first order phase transition for
any finite value of $x$, suggesting again that $x_c = \infty$
\cite{eps}. However, it relies on the hope that $\epsilon = 1$
is small and, therefore, is not conclusive.

2. $x_c= finite$? The absence of a true order parameter for the
gauge-Higgs system is certainly consistent with finite $x_c$.
Moreover, because there is no symmetry breaking, the existence of a
line of second order phase transitions starting at $x_c$ is very
unlikely. However, the proof of the fact that the Higgs and symmetric
phases are continuously connected \cite{fradk} refers to a lattice
system with a finite cutoff and is not applicable to a continuum
system we are interested in.

\begin{figure}[t]
\vspace*{-0cm}
\hspace{-1cm}
\epsfysize=15cm
\centerline{\epsffile{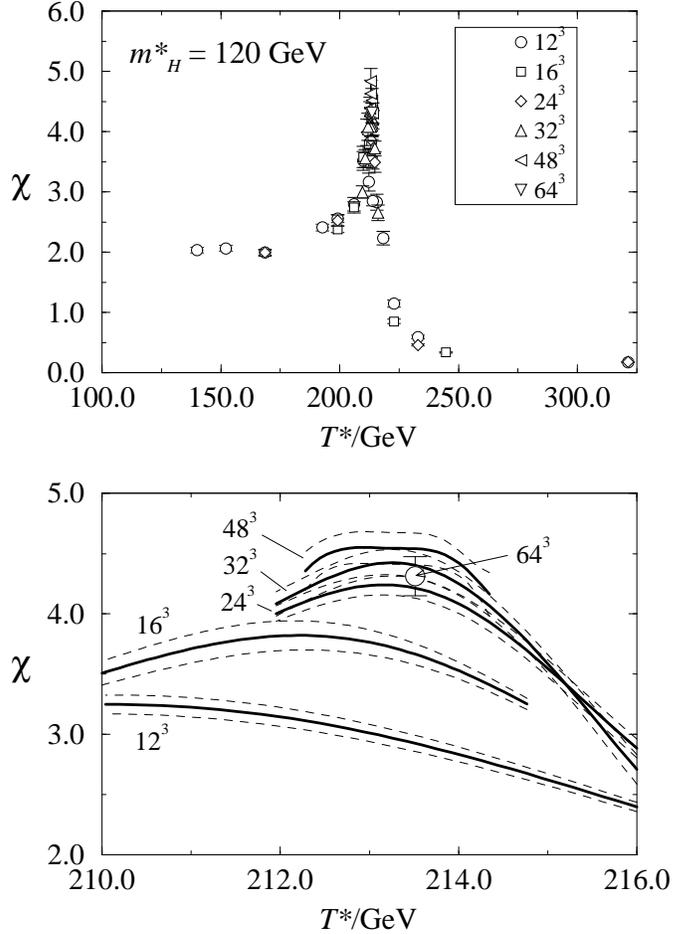}}
\vspace{-2cm}
\caption[a]{\protect $\phi^\dagger\phi$ susceptibility at
$m_H^*=120$~GeV plotted as a function of $T^*$ for lattices of
various sizes. The lower panel shows in more detail the region near
the maximum; the continuous lines with error bands result from
multihistogram reweighting.  The maximum values are plotted in
Fig.~3.}
\end{figure}

A study of one-loop Schwinger-Dyson equations for this system argues
in favour of a finite value of $x_c$ \cite{buch}. However, this
analysis relies heavily on the applicability of perturbation theory
near the phase transition point. This is known to
break down at $m_H \sim m_W$.

In this letter we present strong non-perturbative evidence for the
fact that the line of first order phase transitions has a critical
end-point at a finite value of $x$, $0.09 < x_c < 0.17$, and most
likely $x_c \approx \frac{1}{8}$. In terms of the physical Higgs mass in
the MSM this means that the phase transition ends between $m_H = 70$
and $95$ GeV, probably near $m_H = 80$ GeV.

\begin{figure}[t]
\vspace*{-1.2cm}
\hspace{-1cm}
\epsfysize=16cm
\centerline{\epsffile{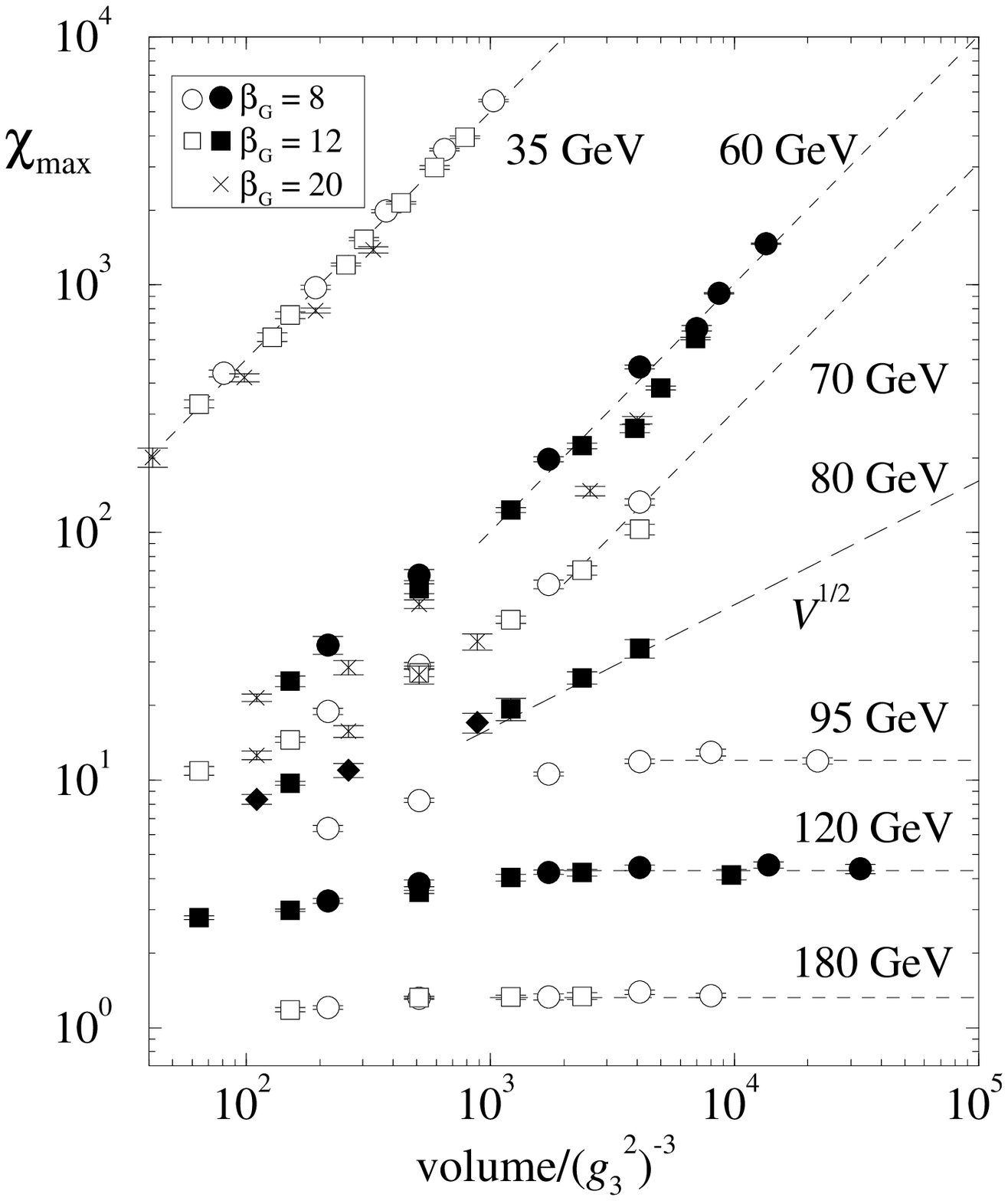}}
\vspace{-3.5cm}
\caption[a]{\protect The maximum values $\chi_{\rm max}$ for
different $m_H^*$ plotted as a function of $V$. The dashed lines
are lines $\sim V,V^{1/2},V^0$.}
\end{figure}

The lattice action corresponding to (\ref{univers}) is, in standard
notation,
\begin{eqnarray}
S&=& \beta_G \sum_x \sum_{i<j}(1-\frac{1}{2} 
{\rm Tr\,} P_{ij}) +\nonumber \\
 &-& \beta_H \sum_x \sum_i
\frac{1}{2}{\rm Tr\,}\Phi^\dagger(\mbox{\bf x})
	U_i(\mbox{\bf x})\Phi(\mbox{\bf x}+i) +
\label{lagrangian} \\
 &+& \sum_x
\frac{1}{2}{\rm Tr\,}\Phi^\dagger(\mbox{\bf x})\Phi(\mbox{\bf x}) 
	+ \beta_R\sum_x
 \bigl[ \frac{1}{2}{\rm Tr\,}\Phi^\dagger(\mbox{\bf x})
\Phi(\mbox{\bf x})-1 \bigr]^2.
\nonumber
\end{eqnarray}
Here $g_3^2a=4/\beta_G, x=\beta_R\beta_G/\beta_H^2$; y is given
in terms of $\beta_H,x,\beta_G$ in \cite{3dlatt}. The continuum
limit $a\to0$ corresponds to $\beta_G\to\infty, \beta_H\to1/3,
\beta_R\to0$.

Among the many tests of the order of the transition we shall use here
(I) the finite size scaling analysis of the $\phi^\dagger\phi$
susceptibility and (II) the analysis of the correlation lengths.  We
define the dimensionless $\phi^\dagger\phi$ susceptibility
\begin{equation}
\chi = g_3^2 V \left\langle (\phi^\dagger\phi
     - \langle \phi^\dagger\phi\rangle)^2\right\rangle
\end{equation}
and measure it as a function of $T^*$.  For each volume we find
the provisional `transition temperature' $T^*_{t,V}$ where $\chi$
attains its maximum value $\chi_{\rm max}$.  There are now 3 distinct
possibilities: a) In a first order phase transition
$\langle\phi^\dagger\phi\rangle$ has a discontinuous jump
$\Delta_\phi$, and $\chi_{\rm max}\propto V\times\Delta_\phi^2$.  b)
In a second order transition $\chi$ displays critical behaviour, and
$\chi_{\rm max} \propto V^\gamma$, where $\gamma$ is a critical
exponent \cite{finitesize}.  c) If there is no transition, $\chi$ is
regular and remains finite when $V\rightarrow\infty$ (on a system
with periodic boundary conditions).

Fig.\,2 shows $\chi(T^*)$ measured from lattices of sizes
$12^3$--$64^3$ for $m^*_H=120$\,GeV and $\beta_G\equiv 4/(g_3^2a)
= 8$.  The data exhibits a strong peak at $T^* \sim 213$\,GeV,
suggesting a possibility of a phase transition.  However, on closer
inspection (bottom panel of Fig.\,2), one sees that $\chi_{\rm max}$
remains finite (within the statistical accuracy), and the provisional
transition is only a sharp -- but regular -- cross-over.  The maximum
values $\chi_{\rm max}$ for different $m_H^*$ are shown as a
function of $V$ in Fig.\,3\@.
Note that the natural unit $g_3^2$ is used in writing
\begin{equation}
V(g_3^2)^3 = ({V/a^3})({4/\beta_G})^3=
({4N/\beta_G})^3.
\end{equation}
In Fig.\,3 we use 3 different lattice spacings ($\beta_G = 8$, 12,
20); no significant finite lattice spacing effect can be observed (the
scatter in $m_H^*=60$\,GeV is due to the large variation in lattice
geometries: some volumes are long cylinders, some cubes).  The pattern
of Fig.\,3 very clearly suggests that the behaviour of the system
changes around $m^*_H = 80$\,GeV from a 1st order transition to no
transition.  The line $\sim V^{1/2}$ corresponds to mean field
critical behaviour.  The data in Fig.\,3 cannot yet distinguish the
true critical exponent, nor whether $m_H^*=80$\,GeV is actually above
or below the critical $m_H^*$: near the critical point increasingly
large volumes are needed in order to see the asymptotic behaviour.

Another evidence of the absence of the phase transition comes from the
study of the correlation lengths of the system. If $x_c = \infty$ then
the phase transition becomes weaker when the Higgs mass is
increased. The jump of the order parameter $\phi^{\dagger}\phi$ gets
smaller together with the mass of the scalar excitation.  At the same
time, the vector correlation length may remain finite at the transition
point, making the resolution of the nature of the phase transition to be
a very difficult problem to solve numerically because of the increasing
hierarchy of the scalar and vector masses.  A typical
signature of this situation is a drastic increase of the scalar
correlation length for all $x$ at some value of $y(x)$.

\begin{figure}
\vspace*{-1cm}
\epsfxsize=10cm
\centerline{\hspace{-1cm}\epsffile{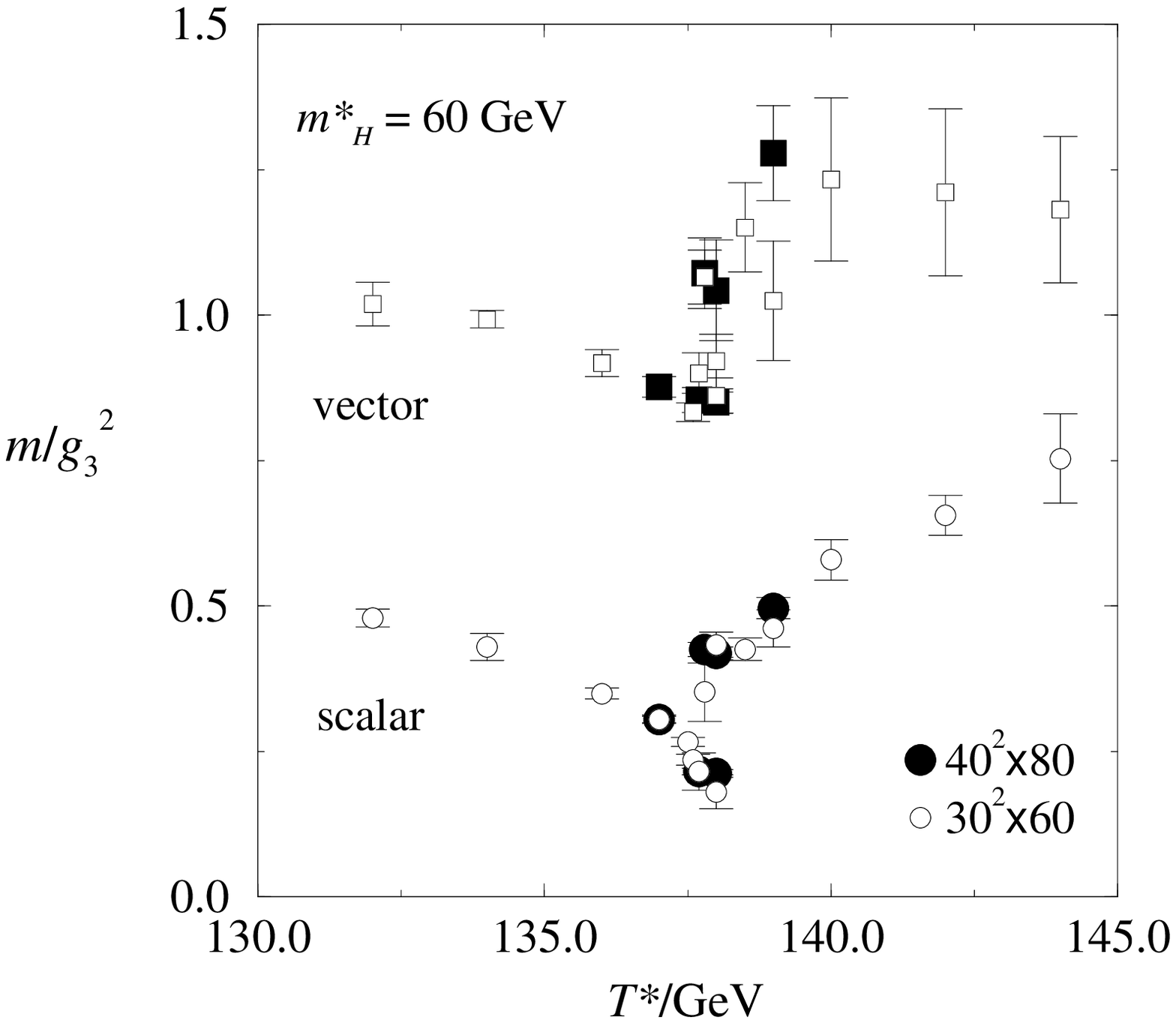}}
\vspace*{-5.7cm}
\epsfxsize=10cm
\centerline{\hspace{-1cm}\epsffile{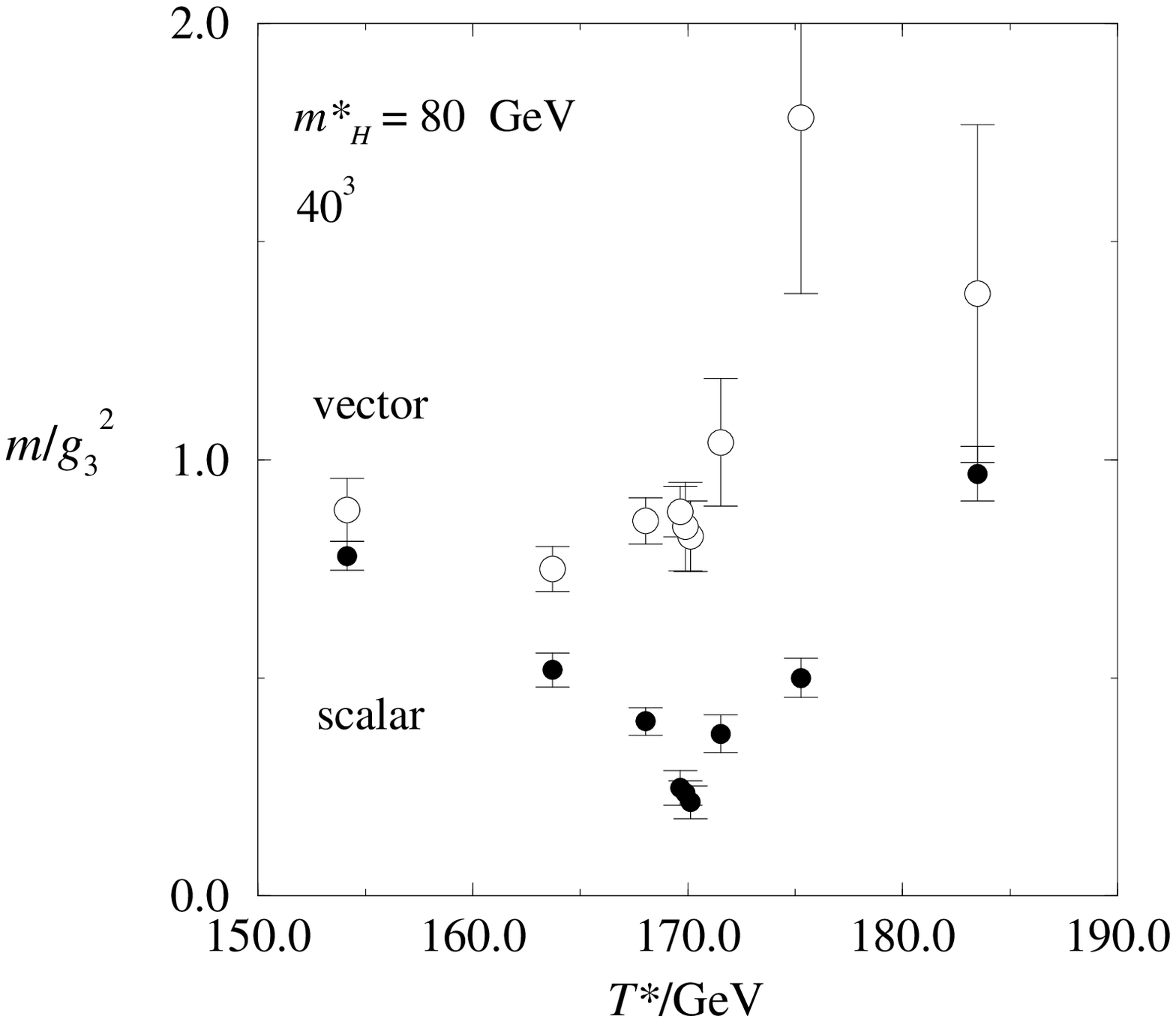}}
\vspace{-4cm}
\caption[a]{The scalar and vector mass dependence on the temperature
for ``small'' Higgs masses, $m_H^*=60$ and $80$ GeV}
\end{figure}

\begin{figure}
\vspace*{-1cm}
\epsfxsize=10cm
\centerline{\hspace{-1cm}\epsffile{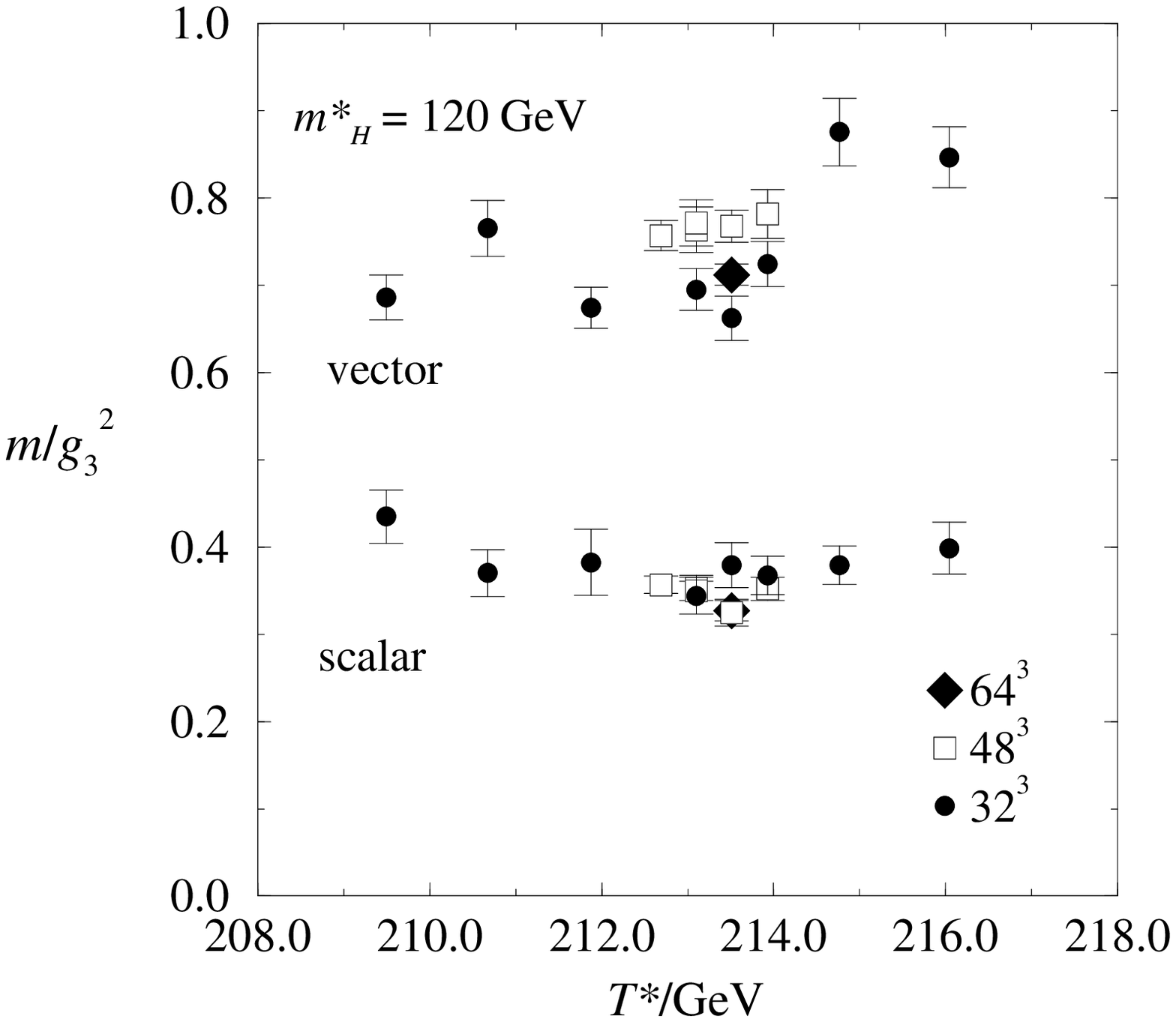}}
\vspace*{-5.7cm}
\epsfxsize=10cm
\centerline{\hspace{-1cm}\epsffile{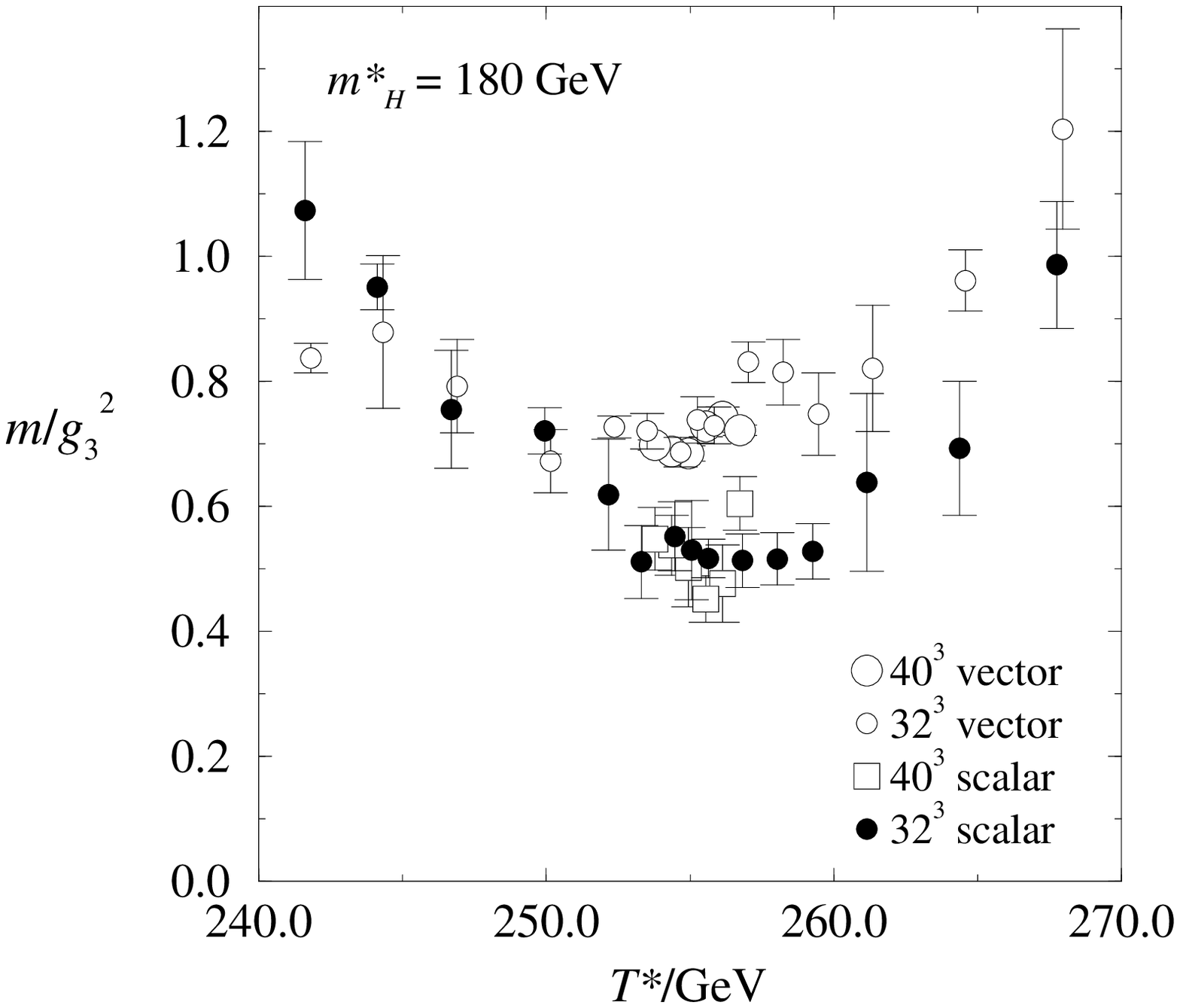}}
\vspace*{-4cm}
\caption[a]{The scalar and vector mass dependence on the temperature
for ``large'' Higgs masses, $m_H^*=120$ and $180$ GeV.}
\end{figure}

If, on the contrary, $x_c$ is finite, then for $x > x_c$ all
correlation lengths of the system are finite, and expectation values
of different gauge-invariant operators are continuous functions of
$y$.  After some minimum size, finite volume effects become
negligible. In this case a reliable lattice MC analysis, which is
hardly possible to carry out near $x_c$, becomes comparatively quite
simple at large Higgs masses.

On Figs.\,4 and 5 we present the behaviour of the scalar and vector
masses (the inverse $\pi$ and $V_j^{0}$ correlation lengths) for
$m_H^* = 60,\,80,\,120$ and $180$\,GeV near the transition/cross-over
temperature.  Fig.\,4a clearly demonstrates the jump of the
correlation lengths typical of 1st order transitions.  Fig.\,4b shows
the power-like decrease of the mass of the scalar excitation with no
change of the vector mass across the critical region.  In contrast,
the behaviour of scalar and vector masses is smooth for $m_H^*=120$
and $180$\,GeV (Fig.\,5), signalling the absence of the phase
transition.  Within the statistical accuracy, the masses and the
susceptibility $\chi$ are independent of the lattice spacing, 
showing that the observed behaviour is not a lattice artefact and
persists in the continuum limit.

To summarize, we demonstrated that the Higgs and confinement phases
of 3d SU(2) gauge-Higgs model can be continuously connected. This
means that the electroweak phase transition in weakly coupled
electroweak theories is absent in a part of their parameter space.
For the minimal standard model the critical value of the Higgs mass
is near $80$ GeV.

\end{document}